%% file: main.tex
\documentclass[runningheads]{llncs}
\newcommand{\equalcontrib}{\textsuperscript{*}}
\usepackage[T1]{fontenc}
\usepackage{lipsum} 
\usepackage{graphicx,verbatim}
\usepackage{multirow}
\usepackage{booktabs}  
\usepackage{amssymb}
\usepackage{graphicx}
\usepackage{array}
\usepackage{multirow}
\usepackage{romannum}
\usepackage{amsmath}
\usepackage{hyperref}
\usepackage{makecell}
\usepackage{booktabs} 
\usepackage{multirow} 
\usepackage[table]{xcolor}  
\usepackage{colortbl}
\usepackage{array} 
\usepackage{makecell}

\usepackage{caption}
\usepackage{color}

\begin{document}
\pagenumbering{arabic}

\newcommand{\OurMethod}{\textit{RetinaLogos}}

\newcommand{\OurData}{\textit{RetinaLogos-1400k}}

\title{\OurMethod: \\Fine-Grained Synthesis of High-Resolution Retinal Images Through Captions}
\titlerunning{\OurMethod}


\author{
Junzhi Ning\equalcontrib$^{1}$, 
Cheng Tang\equalcontrib$^{1,3}$, 
Kaijing Zhou$^{4}$, 
Diping Song$^{1}$, 
Lihao Liu$^{1}$, \\
Ming Hu$^{1,5}$, 
Wei Li$^{6}$, 
Huihui Xu$^{1}$,
Yanzhou Su$^{7}$, 
Tianbin Li$^{1}$,
Jiyao Liu$^{8}$, \\
Jin Ye$^{5,1}$, 
Sheng Zhang$^{9}$, 
Yuanfeng Ji$^{10}$, 
Junjun He$^{1,2}$\textsuperscript{\dag}
}


\institute{
$^{1}$ Shanghai Artificial Intelligence Laboratory, China\\
$^{2}$ Shanghai Innovation Institute, China \\
$^{3}$ Shanghai Institute of Laser Technology, China\\
$^{4}$ Eye Hospital, Wenzhou Medical University, China\\
$^{5}$ Monash University, Australia\\
$^{6}$ Shanghai Jiao Tong University, China\\
$^{7}$ Fuzhou University, China\\
$^{8}$ Fudan University, China\\
$^{9}$ Imperial College London, United Kingdom\\
$^{10}$ Stanford University, USA\\
\email{hejunjun@pjlab.org.cn}
}

\maketitle              
\begingroup
\renewcommand\thefootnote{}
\footnotetext{\equalcontrib~Equal contribution. textsuperscript{\dag} Corresponding author.}
\endgroup

\input{section/0_abstract}

\input{section/1_introduction}

\input{section/2_method}

\input{section/3_experiments}

\input{section/4_conclusion}

\bibliographystyle{splncs04}
\bibliography{mybibliography}

\end{document}

%% file: section/0_abstract.tex
\vspace{-1cm}
\begin{abstract}
The scarcity of high-quality, labelled retinal imaging data, which presents a significant challenge in the development of machine learning models for ophthalmology, hinders progress in the field. Existing methods for synthesising Colour Fundus Photographs (CFPs) largely rely on predefined disease labels, which restricts their ability to generate images that reflect fine-grained anatomical variations, subtle disease stages, and diverse pathological features beyond coarse class categories. To overcome these challenges, we first introduce an innovative pipeline that creates a large-scale, captioned retinal dataset comprising 1.4 million entries, called \OurData. Specifically, \OurData~ uses the visual language model (VLM) to describe retinal conditions and key structures, such as optic disc configuration, vascular distribution, nerve fibre layers, and pathological features. Building on this dataset,   we employ a novel three-step training framework, called \OurMethod, which enables fine-grained semantic control over retinal images and accurately captures different stages of disease progression, subtle anatomical variations, and specific lesion types. Through extensive experiments, our method demonstrates superior performance across multiple datasets, with 62.07\% of text-driven synthetic CFPs indistinguishable from real ones by ophthalmologists. Moreover, the synthetic data improves accuracy by 5\%-10\% in diabetic retinopathy grading and glaucoma detection. Codes are available at \href{https://github.com/uni-medical/retina-text2cfp}{Link}.

\keywords{Retinal Imaging  \and Text-to-Image Generation  \and Medical Data Synthesis \and Fine-grained Controllable generation }

\end{abstract}

%% file: section/1_introduction.tex
\section{Introduction}

Eye healthcare has become a major global concern, as untreated ocular conditions can severely impact an individual’s quality of life \cite{demmin2020visual}. Many people have difficulty accessing ophthalmic resources, particularly in resource-limited areas \cite{sommer2014challenges}. To overcome the challenges posed by limited ophthalmic resources, early detection of eye diseases is crucial, as it enables timely intervention and can help prevent irreversible vision loss \cite{klein1997diabetic}. Among the available diagnostic tools, non-invasive fundus imaging, particularly Color Fundus Photography (CFP), is one of the most widely used and affordable methods in daily clinical practice. Recent advancements in deep learning have significantly transformed the field, enabling the automated analysis of CFP and offering great promise for the early detection of common eye diseases \cite{schmidt2018artificial,li2025ophora}. Current deep-learning techniques rely heavily on large-scale datasets to train various downstream models for CFP. For instance, training foundational CFP models \cite{zhou2023foundation,du2024ret,silva2025foundation} that are competent in zero-shot downstream tasks requires at least a million-level dataset to facilitate model convergence. Despite these advancements, the scarcity of CFP, both in quantity and quality, emphasizes the urgent need for more high-quality data in this domain.

\begin{figure}[!t]
    \centering
    \includegraphics[width=0.9\linewidth]{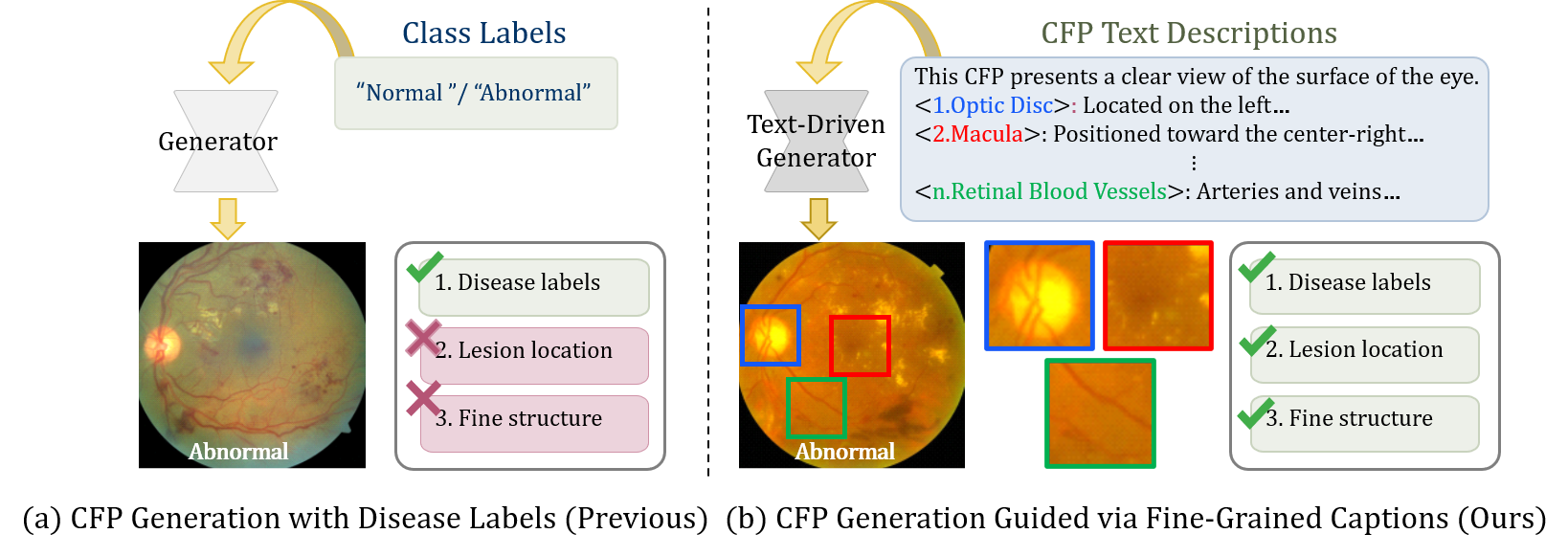}
    \caption{Class-Conditioned CFP Generation vs. Text-Driven CFP Generation.}
    \label{fig:teaser} 
  \end{figure}

Generative models that synthesize data for training various downstream medical tasks have shown significant success \cite{bluethgen2024vision,ning2025unveiling,jiang2020covid,yang2023mammodg,islam2020gan,ning2025unpaired}, and provide a feasible solution to address the issues of data scarcity. For instance, methods in \cite{zhang2024fundus,zhou2020dr,pham2022generating,niu2021explainable,kim2022synthesizing} used Generative Adversarial Networks conditioned on features such as blood vessel structure, lesion region masks, and disease labels to generate retinal images. A two-stage approach has also been adopted in \cite{go2024generation,andreini2021two}, in which the first stage generates realistic conditions, and the second stage generates retinal images based on these conditions. However, current generative methods \cite{shang2023synfundus,go2024generation} primarily rely on the conditions of predefined disease labels, which restrict the generated images to broader categories with diverse anatomical structures. As shown in Fig. \ref{fig:teaser}, this limitation prevents the generation of CFP with more fine-grained details—such as varying stages of retinal disease, subtle anatomical variations, or specific lesion types. 

    To address the above challenges, we first introduce a data collection pipeline designed to amass a large-scale captioned caption dataset totalling 1.4 million real CFPs paired with synthetic detailed captions, which are sourced from both open-source and private datasets. Leveraging this extensive dataset, we then propose \textit{\OurMethod}, a novel text-to-image framework for retinal image synthesis. Specifically, using these extensive text-retinal image pairs, we then develop a tailored text-to-image generator capable of not only synthesizing high-resolution retinal images but also offering fine-grained control over specific anatomical structures and disease progressions. Our method allows for the generation of diverse, visually plausible synthetic CFP, where the appearance can be manipulated through free-form descriptions and prompts.
    
In summary, our main contributions are as follows: a) We propose a comprehensive data collection pipeline, which assembles what is currently the largest \textbf{1.4 million} captioned CFP dataset (1.4 million CFPs paired with synthetic captions) to support advancements in text-driven retinal image synthesis.  b) We propose \textit{\OurMethod}, a novel text-to-image framework for retinal image generation. \textit{To the best of our knowledge, \OurMethod is the first to explore large-scale generation of CFPs from textual descriptions, supported by a dataset exceeding one million CFPs-caption pairs.} c) Our method achieves state-of-the-art performance in text-driven CFP synthesis, demonstrating superior fidelity and clinical relevance on the EyePACS, REFUGE, and IROGS datasets. This has been validated through improved Frechet Inception Distance (FID) and Retina CLIP scores, as well as through evaluations based on criteria defined by expert ophthalmologists.

%% file: section/2_method.tex
\section{Proposed Methodology}
\subsection{Retinal Captioning and Data Synthesis Pipeline}
\begin{figure}[!t]
\setlength{\abovecaptionskip}{0pt}  
\setlength{\belowcaptionskip}{5pt}  
\setlength{\textfloatsep}{0pt}  
\setlength{\intextsep}{0pt}  
    \centering
    \includegraphics[width=0.9\linewidth]{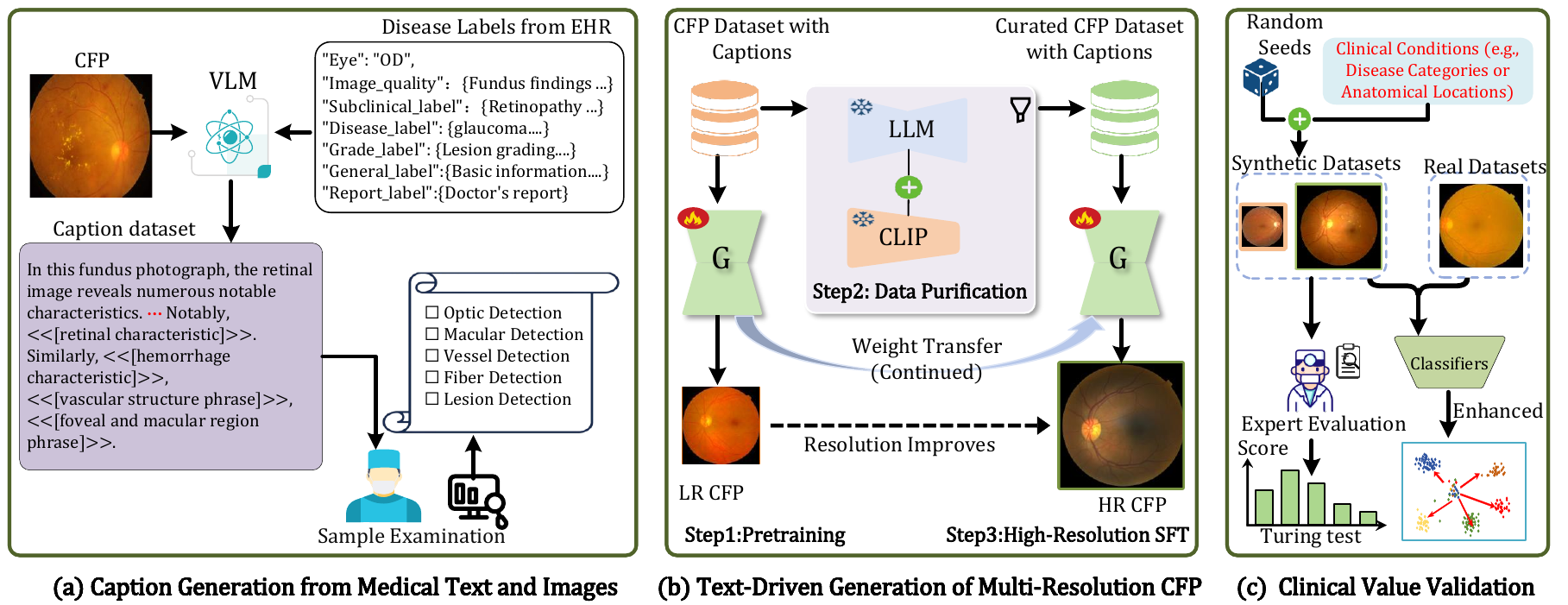
    }
    \caption{\textbf{\textit{\OurMethod} architecture overview.} (a) EHR text and CFP images are integrated via a vision-language model for clinical caption generation. (b) Multi-resolution CFP synthesis includes low-resolution generation, LLM-guided data purification, and high-resolution fine-tuning. (c) Generated CFPs are tested for authenticity, disease classification, and expert evaluation.}
    
    \label{fig:method1}
\end{figure}

\textbf{Authentic Image Quantity $\&$ Diversity.}  
The scale and diversity of the dataset are critical factors influencing the performance of generative models. We constructed a comprehensive dataset of CFPs and corresponding captions, comprising over 1.4 million real-world fundus images sourced from both open-access and private datasets. This dataset includes both images and their corresponding Electronic Health Records (EHRs), as illustrated in Fig. \ref{fig:method1}(a). The fundus images span a broad spectrum of retinal diseases, while the associated EHRs provide essential information to provide grounded labels, including subclinical disease labels, disease severity ratings, and the general health status of patients. Additionally, the EHRs contain diagnostic reports contributed by healthcare professionals. \\
\textbf{Caption Generation $\&$ Reliability.}  
As shown in Fig. \ref{fig:method1}(a), captions are generated using a powerful VLM with the CFP and its corresponding EHR as multimodal inputs. To be more specific, in our proposed data construction pipeline, the VLM is prompted to function as a professional retinal imaging expert, denoted as model $\mathcal{E}$, to generate detailed descriptions based on diagnostic symptomatology extracted from EHR and the corresponding paired fundus images. Let $\{T_i\}_{i=1}^N$ represent the EHR data and $\{D_{i}^l\}_{i=1}^N$ denote the corresponding paired fundus images. The captions generation process can be formally formulated as follows:
\begin{equation}
C^l_{i} = \mathcal{E}(T_i, D_{i}^l), \quad i = 1, 2, \ldots, N.
\end{equation}
Furthermore, to ensure that the captions align with clinical expectations, professional ophthalmologists are involved in reviewing the generated annotations, represented as paired text-to-image data $\{ D_{i}^l, C_{i}^l\}_{i=1}^N$.

\subsection{Retinal Image Synthesis via Text-to-Image Generation Framework}
In this study, we trained a latent flow-matching DIT model inspired by \cite{gao2024lumina}, as our retinal text-to-image generator. We employed the frozen Google Gemma 2B \cite{team2024gemma} text decoder to obtain the word embeddings of the retinal image captions and leveraged the flow matching mechanism to linearly interpolate between noise and the clean sample. Mathematically, given an CFP sample $x^{*} \sim p_{data}$, an associated caption $\phi$, and $\epsilon \sim \mathcal{N}(0, I)$, the linear interpolation forward process is formulated as $x_t=\alpha_t x^*+\beta_t \epsilon = tx^* + (1-t) \epsilon$, where $t \in [0,1]$. Its corresponding vector field is $v_{t}(x_t) = x^* - \epsilon$. During training, the model is optimized using the following conditional flow-matching objective: 
\begin{equation}
    \mathcal{L}_{\text{CFM}} = \mathbb{E}_{t \sim U(0,1), x^* \sim p_{data}, \epsilon \sim \mathcal{N}(0, \mathbf{I})} \left[\left\|v_\theta\left(x_t, t, \phi\right) - v_{t}(x_t)\right\|_2^2\right]
\end{equation}
The training process consisted of three steps, as outlined in Fig.\ref{fig:method1}(b).\\

\noindent \textbf{Step \Romannum{1}: Pretraining Stage.} 
The pretraining stage initialised the model weights using the checkpoint from \cite{zhuo2024lumina} as a starting point, the warm-up approach reduced training time by closing the gap between the visual representations of natural and retinal images. The primary objective of this stage was to effectively adapt the model’s backbone for the retinal generation task. The model was pre-trained on 1.4 million retinal images, where the corresponding annotated descriptions are denoted as $\{C_{i}^l\}_{i=1}^N$, at a low resolution of $256 \times 256$, with the corresponding images represented as $\{D_{i}^l\}_{i=1}^N$. 

\noindent \textbf{Step \Romannum{2}: Precision Filtering and Semantic Refinement.}
Ensuring data quality is indispensable for training text-driven image generation models for CFPs, we implement two strategies to further purify the datasets of image-text pairs. Firstly, we employ the existing RetinClip encoders \cite{silva2025foundation} to sift through retinal images and caption annotations, filtering out pairs with CLIP similarity scores below 0.6, an operation captured by:
\begin{equation}
\{D_{i}^h, C_{i}^h\}_{i=1}^{N} = \{ (D_{i}^l, R(C_{i}^l)) \mid S(D_{i}^l, C_{i}^l) \geq 0.6~s.t.~i\in \{1,2,3...,N\}\},
\end{equation}
where \(S(D_{i}^l, C_{i}^l)\) quantifies the semantic alignment between an image \(D_{i}^l\) and a caption \(C_{i}^l\), and \(R(C_{i}^l)\) denotes the refined caption. The dataset before filtering and refinement is denoted as \(\{D_{i}^h, C_{i}^h\}_{i=1}^{N}\), which includes all collected pairs  retinal images \(D_{i}^l\) and captions \(C_{i}^l\). After filtering, the resulting dataset \(\{D_{i}^h, C_{i}^h\}_{i=1}^{N}\) contains only the image-caption pairs with sufficient semantic alignment. Then, we further refined the captions with the Qwen 2.5 LLM \cite{yang2024qwen2} using designed medical prompts. The prompt is carefully tailored to eliminate unnecessary descriptions prevalent in the retinal image content (such as recommendations to avoid liability) while preserving the original meaning of the captions.\\
\textbf{Step \Romannum{3}: High-Resolution Supervised Fine-Tuning.} The \OurMethod~ was fine-tuned on higher resolutions, reaching up to 1024×1024, using a dynamic padding strategy from the Next-DIT architecture to enable training with diverse aspect ratios. This allowed the second stage of supervised fine-tuning (SFT) to achieve model convergence more efficiently, even with a relatively limited dataset compared to the scale of natural images. This high-resolution training enhanced image detail, allowing the model to capture finer retinal features.

\subsection{Evaluation Standard for Generated Text-to-Retinal Images}

As shown in Fig.~\ref{fig:method1}(c), assessing the quality of synthesized retinal images derived from the provided captions is key to their clinical relevance. We employ both downstream task validation and expert evaluation by medical professionals. In particular, in collaboration with ophthalmologists, we designed the first evaluation principles based on five key anatomical structures—the optic disc, macula, retinal vasculature, retinal nerve fibre layer, and pathological lesions—to measure the quality of CFPs generated through a text-driven synthesis approach.

%% file: section/3_experiments.tex
\section{Experiments and Discussion}

\subsection{Dataset and Training Details}

\begin{table}[!tb]
\centering
\caption{\textbf{Quantitative Comparison of Synthetic CFP with Real CFP.} The FID and KID metrics assess the similarity between generated and real images, while the Inception Score (IS) measures the diversity of generated images. \textsuperscript{$\dagger$} Results for Lumina-Next are based on weights pre-trained without medical retinal image data. * EyePACs and AIROGS originate from the same institution.}
\label{tab:results}
\resizebox{\linewidth}{!}{%
\begin{tabular}{l|ccc|ccc|c} 
\toprule\toprule
\multirow{2}{*}{\textbf{Dataset}} & \multicolumn{3}{c|}{\textbf{FID}$\downarrow$} & \multicolumn{3}{c|}{\textbf{KID}$\downarrow$} & \multirow{2}{*}{\textbf{Inception Score}} \\ 
\cmidrule(lr){2-4} \cmidrule(lr){5-7} 
             & APTOS & EyePACs & AIROGS & APTOS & EyePACs & AIROGS \\ \midrule
APTOS\cite{karthik2019aptos}                  &  -     &    52.931     &    42.904     &    -    &    0.0417~(0.0015)   &    0.0342~(0.0015)    &   1.969       \\
EyePACs\cite{gulshan2016development}                 &  52.931    &    -    &    *11.005     &    0.0417~(0.0016)   &    -    &    *0.0066~(0.0008)  &  2.132         \\
AIRROGS\cite{de2023airogs}               &  42.905    &   *11.005      &   -    &  0.0342~(0.0015)   &    *0.0066~(0.0008)     &    -    &   1.993     \\ \midrule
\textbf{Average}         &  47.918   &   52.931   & 42.904 &  0.03795    &   0.0417   &  0.0342    & 2.031 \\ \Xhline{2\arrayrulewidth}
MedFusion\cite{muller2023multimodal}             &   77.022    &   68.162      &    60.651 &    0.0871~(0.0016)    &    0.0748~(0.0013)    &    0.0716~(0.0015)   &   1.828     \\  
Lumina-Next\textsuperscript{$\dagger$}\cite{zhuo2024lumina} &    240.406     &    247.135  &   251.953    &    0.1800~(0.0024)    &    0.1830~(0.0023)    &    0.1938~(0.0026)   &   7.615             \\ \midrule
  \textbf{Ours}          &    \textbf{56.078}   &    \textbf{42.437}     &   \textbf{35.190}      &    0.0369~(0.0012)    &   0.0230~(0.0008)    &   0.021~(0.0007)   &    1.864      \\
\bottomrule\bottomrule
\end{tabular}%
}
\end{table}

\noindent \textbf{Datasets.} We evaluate the authenticity of generated CFP using three benchmark datasets: APTOS\cite{karthik2019aptos}, EyePACs\cite{gulshan2016development} and AIROGS\cite{de2023airogs}. Additionally, we assess the performance of our synthetic data in downstream classification tasks using the IDRiD\cite{porwal2018indian} and REFUGE2\cite{fang2022refuge2} datasets. These datasets contain 5,000, 35,126, and 113,893 color fundus images, respectively. The IDRiD dataset includes 516 fundus images. The REFUGE2 dataset consists of 1,200 fundus images, of which we selected 800 labelled images for use in our experiments.

\noindent \textbf{Implementation Details.} Our framework is implemented in PyTorch and trained on 8 Nvidia RTX A100 GPUs, we trained \OurMethod~with total iterations of 1M with the learning rate of $1 \times 10^{-5}$. For the authenticity evaluation, we compared the generated CFPs with real CFPs from the existing open-sourced datasets using metrics of FID, Kernel Inception Distance(KID), and Inception Score(IS). To quantify the text-to-image alignment, we leveraged the existing foundational CLIP-based method \cite{silva2025foundation} to measure the caption-to-image similarity score. In downstream classification tasks,  we trained models on different datasets for 100 epochs. ResNet-50 processed images at a resolution of 512×512 pixels, whereas ViT-B/16 operated on 224×224 pixels.

\begin{figure}[!tb]
    \centering
    \includegraphics[width=0.9\linewidth]{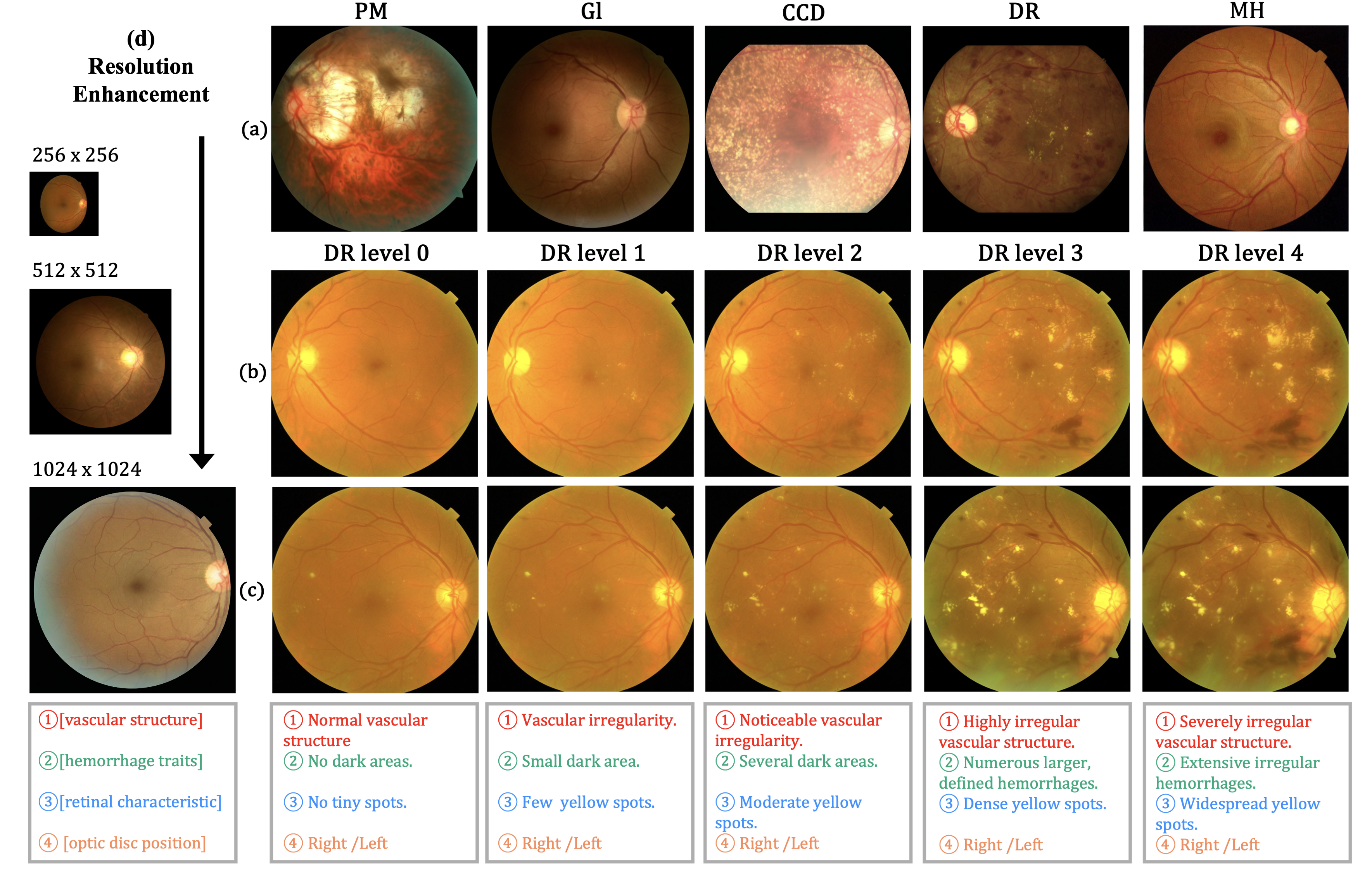}
    \caption{\textbf{Visual Comparison of Generated CFPs under Different Stages, Resolutions and Pathological Structures.} 
    (a) Disease Categories: Displays different retinal disease types, including pathological myopia (PM), glaucoma (GI), crystalline corneoretinal dystrophy (CCD), diabetic retinopathy (DR), and macular hole (MH). 
    (b) \& (c) Diabetic Retinopathy (DR) Levels: Demonstrates the progression of DR across severity levels (0--4).
    (d) Resolution Enhancement up to 1024 progressively.}
    \label{fig:Comparsion}
\end{figure}

\begin{table}[!tb]
\centering
\caption{\textbf{Performance of Diabetic Retinopathy Grading Classification and Glaucoma Detection.} Results for real and synthetic datasets using \OurMethod, with the proposed synthetic CFP data (+ours). Metrics include Accuracy (Acc), F1-Score, and Quadratic Weighted Kappa (QWK).}
\label{tab:downstream}
\resizebox{0.9\linewidth}{!}{%
\begin{tabular}{ccccc|ccc} 
\toprule \toprule
\multirow{2}{*}{\textbf{Training Set}} & \multirow{2}{*}{\textbf{Extra Data}} & \multirow{2}{*}{\textbf{\#Samples}} & \multirow{2}{*}{\textbf{Eval Set}}  & \multirow{2}{*}{\textbf{Model}} & \multicolumn{3}{c}{\textbf{Metrics}} \\
\cmidrule(lr){6-8}  
 & & & & &  \textbf{Acc} & \textbf{F1-Score} & \textbf{QWK} \\
\midrule
\multirow{4}{*}{IDRiD-Train} & N/A   & 413      & \multirow{4}{*}{\centering IDRiD-Eval} & ResNet-50 & 0.5436 & 0.4598  & 0.6223 \\
                           & N/A   & 413      &                                        & ViT-B/16 & 0.4563 & 0.3589  & 0.3848 \\
                           & +ours & 9837+413 &                                        & ResNet-50 & 0.6375 & 0.5183  & 0.6868 \\
                           & +ours & 9837+413 &                                        & ViT-B/16 & 0.5533 & 0.5400  & 0.6213 \\
\midrule
\multirow{4}{*}{REFUGE2-Train} & N/A   & 640      & \multirow{4}{*}{\centering REFUGE2-Test} & ResNet-50 & 0.8562 & 0.7842 & 0.3935 \\
                              & N/A   & 640      &                                          & ViT-B/16  & 0.9037 & 0.8615 & 0.6833 \\
                              & +ours & 9117+640 &                                          & ResNet-50 & 0.9375 & 0.8537 & 0.5505 \\
                              & +ours & 9117+640 &                                          & ViT-B/16  & 0.9762 & 0.9313 & 0.8627 \\
\bottomrule \bottomrule
\end{tabular}
}

\end{table}
 
\subsection{Experimental Results}
\noindent \textbf{Comparison Results on Authenticity and Classification Tasks.}  
As shown in Table \ref{tab:results}, we compare our method with MedFusion\cite{muller2023multimodal} and Lumina-Next\cite{zhuo2024lumina}. Our method achieves results closest to the real image benchmarks. Visual comparisons are presented in Fig. \ref{fig:Comparsion} (a), where CFPs for different eye diseases are generated based on text descriptions. Fig. \ref{fig:Comparsion} (b) and (c) demonstrate the controllable generation of Diabetic Retinopathy progression in the left and right eyes under fixed random seeds, achieved by varying the descriptions of anatomical and pathological symptoms. In downstream tasks, we focused on exploring the classification performance of ophthalmological diseases using our augmented training dataset with synthetic CFPs, as detailed in Table \ref{tab:downstream}. Our augmented training data consistently enhances disease classification performance regardless of backbones, leading to an accuracy increase of 5 \%–10\%. Fig.\ref{fig:Comparsion} (d) demonstrates the ability of our RetinaLogos to scale up resolution levels, enabling high-quality CFP generation.

\begin{table}[!tb]
    \centering
    \caption{\textbf{Clinical Evaluation of Generated Colour Fundus Photographs (CFPs).} Authenticity is assessed based on the predictive outcomes for real and synthetic CFPs, including those with no pathological symptoms. The expert evaluation considers the clinical assessment based on clinician-proposed criteria. Values in bold indicate superior performance. The scale of the expert evaluation is scored from 0 to 3 to reflect the level of semantic resemblance between the caption and the generated CFP. }
    \label{tab:combined_evaluation}
    \resizebox{0.9\linewidth}{!}{%
    \begin{tabular}{ll|cc}
    \toprule\toprule
    \multirow{2}{*}{\textbf{Test Categories}} & \multirow{2}{*}{\textbf{Evaluation Aspects}} & \multicolumn{2}{c}{\textbf{Prediction Outcome}} \\
    \cmidrule(lr){3-4}
     &  & \textbf{Real Image} & \textbf{Synthetic Image} \\
    \midrule
    \multirow{2}{*}{Authenticity} 
        & \textit{Real Image } & \textbf{64.29\%} & 35.71\% \\
        & \textit{Synthetic Image } & \textbf{62.07\%} & 37.98\% \\
    \midrule
    \multirow{6}{*}{Retinal Evaluation}
      & \textit{Optic Disc Structure and Position} & 2.11 & \textbf{2.41} \\
      & \textit{Macular Structure and Position} & 2.10 & \textbf{2.45} \\
      & \textit{Vascular Structure and Position} & 2.21 & \textbf{2.55} \\
      & \textit{Retinal Nerve Fiber Layer Structure} & \textbf{2.46} & 2.41 \\
      & \textit{Lesion Structure and Position} & 1.46 & \textbf{1.90} \\
      & \textit{Overall CFP Image Quality} & 2.07 & \textbf{2.28} \\
    \midrule
      & \textbf{Average} & 2.06 & \textbf{2.33} \\
    \bottomrule \bottomrule
    \end{tabular}
    }
\end{table}

\begin{figure}[!tb]
\centering
\begin{minipage}{0.47\textwidth} 
\centering
\includegraphics[width=0.6\linewidth]{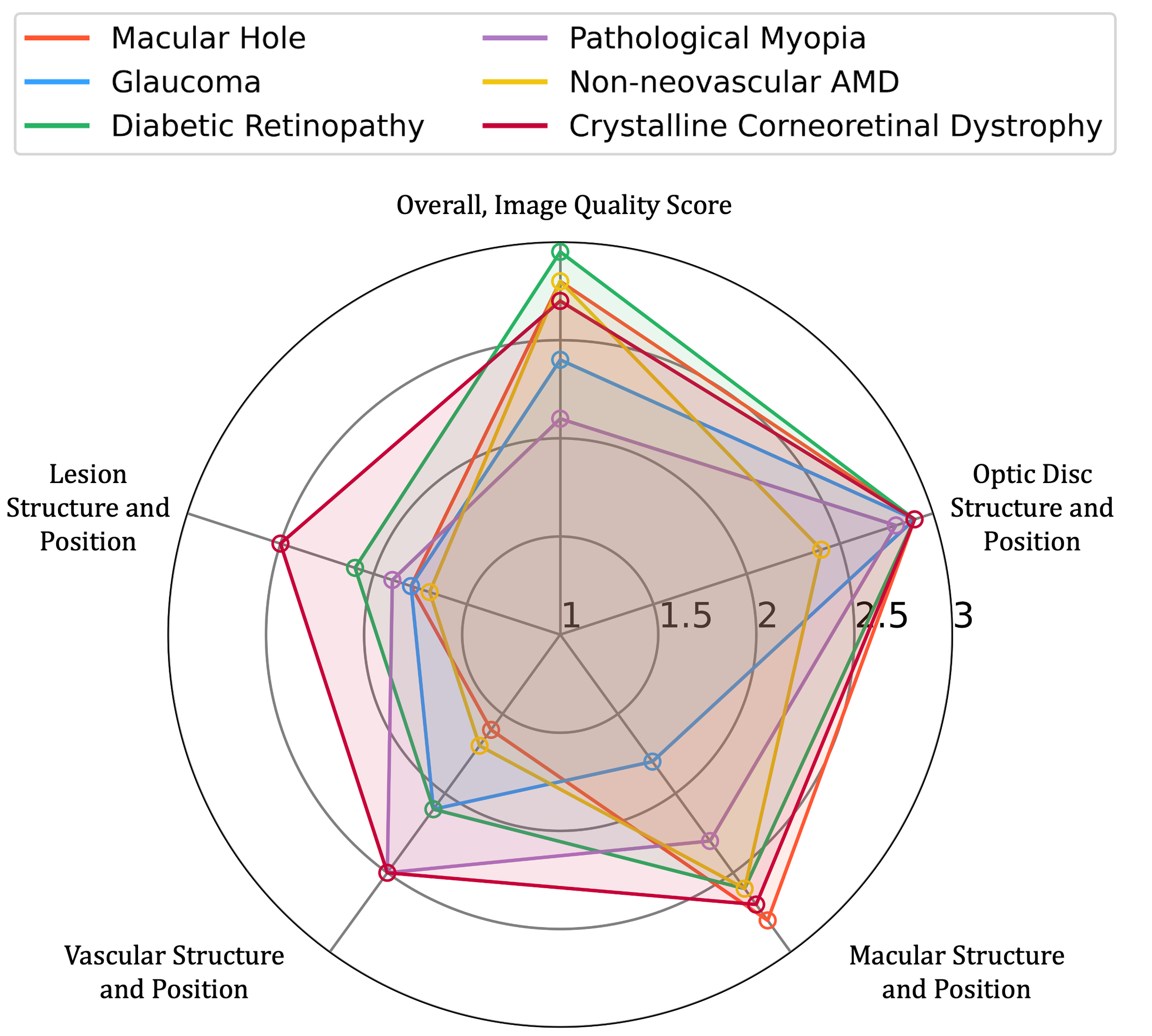} 
\caption{Comparison of ophthalmologists’ evaluation scores on individual retinal disease criteria.}
\label{fig:radar_chart}
\end{minipage}%
\hfill 
\begin{minipage}{0.5\textwidth} 
\centering
\captionsetup{type=table} 

\caption{\textbf{Ablation Studies for Generators Configurations and Components.} Abbreviations: PT = Pre-train, PL = Prolonged Training, HR = High-Resolution, SR = Clip-Selection \& Caption Refinement.}
\label{tab:ablation}
\resizebox{1\linewidth}{!}{%
\begin{tabular}{c|cccc|ccc|c}
\toprule\toprule
\multirow{2}{*}{\textbf{Exp}} & \multicolumn{4}{c|}{\textbf{Components}} & \multicolumn{3}{c|}{\textbf{FID} $\downarrow$} & \multirow{2}{*}{\textbf{Clip Score} $\uparrow$} \\ \cline{2-8} 
& PT & PL & HR & SR & APTOS & EyePACs & AIROGS&   \\ \hline
\Romannum{0}                    &  &       &     &              & 240.406 & 247.135 & 251.953 & 0.5398 \\
\Romannum{1}                    & $\checkmark$ &       &     &              & 63.056  &  75.386  &    64.696     &    0.5485    \\
\Romannum{2}                    & $\checkmark$ & $\checkmark$ &                &   &  72.389     & 71.766  &  63.866  &  0.5439   \\
\Romannum{3}                    & $\checkmark$ & $\checkmark$ &                & $\checkmark$  &  68.786    & 73.635 & 61.465 &  0.5730   \\
\Romannum{4}                    & $\checkmark$ & $\checkmark$ & $\checkmark$ &              &  66.525    &   69.824   &  61.718             &  0.5396  \\ \midrule
\Romannum{5}                    & $\checkmark$ & $\checkmark$ & $\checkmark$ & $\checkmark$   & 56.078  &    42.437  &   35.190  &   0.6601  \\ 

\bottomrule\bottomrule
\end{tabular}%
} 
\end{minipage}
\end{figure}

\noindent\textbf{Ablation Study.} Table~\ref{tab:ablation} summarizes five ablation settings. Exp~\Romannum{1} (\emph{PT}) serves as the $256\times256$ baseline. Exp~\Romannum{2} introduces prolonged training (\emph{PL}), but the marginal improvement suggests that simply extending training does not significantly enhance performance. Exp~\Romannum{4} applies higher resolution (\emph{HR}), indicating that scaling alone without data refinement offers limited benefit. Exp~\Romannum{3} incorporates the caption-refinement module (\emph{SR}), highlighting the importance of data quality. Finally, Exp~\Romannum{5} combines all components, achieving the best overall performance.

\noindent \textbf{Expert Evaluation.} To evaluate the generator's ability to produce clinically relevant CFPs, we conducted authenticity tests by comparing the generated images with real ones and performed an expert evaluation focusing on five key aspects. As shown in the Table. \ref{tab:combined_evaluation}, 62.07\% of the generated CFPs were classified as real, which indicates that the model is capable of producing high-quality images that closely resemble real clinical data. Additionally, Fig. \ref{fig:radar_chart} presents an analysis of the evaluation scores for individual eye diseases based on the generated CFPs.

%% file: section/4_conclusion.tex
\section{Conclusion}
Our work presents \OurMethod, a text-to-image framework that leverages large-scale synthetic retinal caption datasets—comprising 1.4 million entries—to generate high-resolution, clinically relevant retinal images. This approach, which transforms detailed text descriptions into visually rich images capturing key retinal features, has been validated through extensive experiments. Although the controlled generation of the pathological and anatomical structure still leaves room for improvement, particularly in retinal diseases, it shows promising potential to generate CFPs with fine-grained text descriptions in ophthalmology. \newline

\noindent \textbf{Acknowledgments.}
This work was supported by the National Key R\&D Program of China (2022ZD0160101, 2022ZD0160102), the National Natural Science Foundation of China (Grant No.62272450) and Shanghai Artificial Intelligence Laboratory.  \newline

\noindent  \textbf{Disclosure of Interests.} The authors have no competing interests to declare that are relevant to the content of this article.